\documentclass[letterpaper,twocolumn,english,showpacs]{iopart}
\usepackage[T1]{fontenc}
\usepackage[latin9]{inputenc}
\usepackage{babel}
\usepackage{verbatim}
\usepackage{amstext}
\usepackage{amssymb}
\usepackage{graphicx}
\usepackage{esint}
\usepackage[unicode=true,pdfusetitle,
 bookmarks=true,bookmarksnumbered=false,bookmarksopen=false,
 breaklinks=false,pdfborder={0 0 1},backref=false,colorlinks=false]
 {hyperref}
\usepackage{breakurl}

\makeatletter

\usepackage{iopams}
\usepackage{setstack}

\usepackage{babel}

\usepackage[numbers, sort&compress]{natbib}

\def\newblock{\hskip .11em plus .33em minus .07em}

\makeatother

\begin{document}

\title{Easy-axis ferromagnetic chain on a metallic surface}

\author{Alejandro M. Lobos}

\address{Condensed Matter Theory Center and Joint Quantum Institute, Department
of Physics, University of Maryland, College Park, Maryland 20742-4111,
USA.}

\ead{alobos@umd.edu}

\author{Miguel A. Cazalilla}

\address{Graphene Research Centre National University of Singapore, 6 Science
Drive 2, Singapore 117546}

\address{Centro de F{\'i}sica de Materiales CSIC-UPV/EHU and Donostia International
Physics Center (DIPC). Paseo Manuel de Lardizabal, E-20018 San Sebastian,
Spain}

\ead{miguel.cazalilla@gmail.com}
\begin{abstract}
The phases and excitation spectrum of an easy-axis ferromagnetic chain
of $S=1/2$ magnetic impurities built on the top of a clean metallic
surface are studied. As a function of the (Kondo) coupling to the
metallic surface and at low temperatures, the spin chain exhibits
a quantum phase transition from an Ising ferromagnetic phase with
long-range order to a paramagnetic phase where quantum fluctuations
destroy the magnetic order. In the paramagnetic phase, the system
consists of a chain of Kondo-singlets where the impurities are completely
screened by the metallic host. In the ferromagnetic phase, the excitations
above the Ising gap are damped magnons, with a finite lifetime arising
due to the coupling to the substrate. We discuss the experimental
consequences of our results to spin-polarized electron energy loss
spectroscopy (SPEELS), and we finally analyze possible extensions
to spin chains with $S>1/2$. 
\end{abstract}


\maketitle

\section{\label{sec:intro}Introduction}

Recent progress in scanning tunneling microscopy (STM) and other surface-manipulation
techniques at the atomic scale has made it possible to build low-dimensional
magnetic structures with unprecedented control \cite{Wiesendanger09_Spins_on_surfaces_review,Carbone11_Self-assembled_magnetic_networks_on_surfaces}.
Atomic-scale magnetic structures possess an enormous potential for
applications in spintronics as well as in classical and quantum computation
due to their ability to store and process (quantum) information in
the atomic spins. However, in order to build reliable atomic-size
memories or qubits out of such systems, it is crucial to understand
how the coupling to the environment affects their properties. For
instance, it is known that quantum nanomagnets coupled to electronic
degrees of freedom are prone to the effects of decoherence induced
by Ohmic quantum dissipation \cite{Prokofev00_Theory_of_spin_bath,Weiss99_Quantum_Dissipative_Systems}.%

In addition to their interest for information technologies, low-dimensional
magnetic systems coupled to dissipative environments are also very
important from the point of view of fundamental physics. This is because
they allow to study the crossover from quantum to classical behavior
as the coupling to the environment is increased \cite{leggett_two_state,Weiss99_Quantum_Dissipative_Systems}.
Current theoretical models that describe local quantum degrees of
freedom coupled to sources of dissipation predict rich and interesting
physical phenomena at low temperatures, including quantum phase transitions
and quantum criticality \cite{caldeira&leggett81,chakravarty82_macroscopic_quantum_tunneling,bray82_macroscopic_quantum_tunneling,schmid_instanton},
as well as broken continuous symmetries and long-range order (LRO)
in low-dimensional systems \cite{castroneto97_open_luttinger_liquids,Pankov04_NFL_behavior_and_2D_AFM_fluctuations,Werner05_Dissipative_quantum_Ising_model_in_1D,Werner05_Quantum_Spin_Chains_with_site_dissipation,Orth08_Dissipative_quantum_Ising_model_in_cold_atoms,Lobos12_Dissipative_XY_chain}.
The interplay between quantum fluctuations, which are induced by quantum
confinement and reduced dimensionality, and quantum dissipation, caused
by the interaction with the environment, is at the core of this exotic
behavior. From this perspective, magnetic nanostructures built on
the top of clean metals offer a unique platform to test our understanding
on dissipative quantum systems. For instance, the simplest zero-dimensional
(0D) case of a single magnetic impurity coupled to a metallic environment
{[}such like a Co or Mn adatom sitting on a Au(111) surface{]} is
a practical realization of the celebrated Kondo effect, one of the
most paradigmatic phenomena in the physics of strongly correlated
electrons \cite{hewson}. This phenomenon is the spin-compensation
by the conduction electrons of the magnetic moment $\mathbf{S}_{i}$
of an impurity embedded in a metallic host at temperatures $T\ll T_{K}$,
where $T_{K}$ is the so-called Kondo temperature. Eventually, at
$T=0$, the ground state of the system is a many-body singlet (termed
a `Kondo singlet'), formed by a linear superposition of $\mathbf{S}_{i}$
and the spin of a cloud of conduction electrons that is anti-ferromagnetically
correlated with the impurity local moment. The original magnetic impurity
$\mathbf{S}_{i}$ is thus completely screened by the metallic environment.
In the context of magnetic adatoms deposited on clean metallic surfaces,
this is a well-established phenomenon, which reveals itself as a narrow
Fano resonance at the Fermi energy in the STS spectra \cite{Li98_Kondo_effect_on_single_adatoms,Madhavan98_Tunneling_into_single_Kondo_adatom,knorr02}.

Unfortunately, the extension of the single Kondo impurity to the case
of many interacting magnetic impurities embedded in a metallic host
represents a formidable theoretical challenge. Due to the many-body
nature of the problem, exact methods are only limited to simple cases
(i.e., consisting of few magnetic impurities), and at the cost expensive
numerical calculations \cite{Bulla08_NRG_review}. On the other hand,
dynamical mean-field theory (DMFT) methods \cite{georges_dmft} provide
reliable results in the case of bulk 3D compounds, but are much less
successful when applied to low-dimensional systems.

One particular class of such low-dimensional systems, namely, one-dimensional
(1D) chains of magnetic atoms or organic molecules with active magnetic
atoms, are currently under intensive experimental investigation \cite{Gambardella02_Ferromagnetism_in_1D_quantum_spin_chains,Hirjibehedin06_Mn_spin_chains,DiLullo12_Molecular_Kondo_chain,Loth12_Bistability_in_atomic_scale_AFM,Muelleger12_Kondo_state_and_Kondo_resonance_in_spin_chains}.
Thus, by depositing $\simeq0.1$ monolayer (ML) of Co on Pt(997),
which is a vicinal surface containing steps, chains of magnetic atoms
have been recently created~\cite{Gambardella02_Ferromagnetism_in_1D_quantum_spin_chains}.
The magnetic Co atoms decorate the steps and form ferromagnetic chains,
which, when probed by an external magnetic field, typically exhibit
superparamagnetism. However, in the case of Co chains at temperatures
$T\lesssim8$ K, the magneto-crystalline anisotropy results in relaxation
times which are longer than the experimental typical timescale, and
the response of the system becomes effectively ferromagnetic \cite{Gambardella02_Ferromagnetism_in_1D_quantum_spin_chains}.
Another recently reported system, consisting of chains of Co atoms
on Cu(775), has been shown to undergo a Peierls distortion leading
to dimerization \cite{Zaki2012_Dimerized_spin_chain}. This dimerized
phase has been explained by the strong correlations existing in the
partially filled $d$ shells of the Co atoms, which are in the high
spin configuration. Maintaining the high spin in a geometry where
the overlap between the d orbitals is small and correlations are strong,
favors a ferromagnetic ground state, which in turn makes the charge-density
wave instability possible.

In the above-mentioned systems, the coupling to the metallic substrate
was assumed to be weak, and therefore was neglected in their analysis.
However, as we show below, the metallic environment can have dramatic
effects for the phase diagram at low temperatures, and it cannot be
neglected when studying the magnetic excitations of the chain. With
these systems in mind, in this contribution we focus on the analysis
of an easy-axis ferromagnetic (FM) 1D spin chain deposited on the
top of a metallic surface. Interestingly, as a function of the coupling
with the substrate, we predict a quantum phase transition from an
Ising ferromagnetic phase with long range order to a paramagnetic
phase where quantum fluctuations proliferate and destroy the ordered
phase. Although modifying the coupling to the substrate in the experimental
systems created so far may be hard, in the future other systems can
be created where the coupling to the substrate could be tuned, 
uncovering even more interesting phenomena.

The article is organized as follows: In the next Section, we introduce
a simple model to describe an easy-axis $S=1/2$ ferromagnetic chain
on a metallic substrate. This model is studied in the framework of
mean-field theory in Section~\ref{sec:mft}, and the quantum phase
diagram is thus obtained. The phase diagram is displayed in Fig.~\ref{fig:mf_phase_diag},
and it is one of the most important results of this work. In Section~\ref{sec:holtsein}
we consider the excitation spectra of the ferromagnetic phase. We
show that a weak coupling with the metallic substrate can have important
consequences for the magnetic excitations of the ferromagnetic chain.
This properties of the excitations, and in particular, the substrate-induced
damping, can be accessible by experimental probes such as spin-polarized
electron energy loss spectroscopy (SPEELS). In Fig.~\ref{fig:spin_response},
we plot the spin-response of the 1D spin chain due the substrate-modified
magnon excitation spectrum, which is also an important result of this
work. Finally, in Section~\ref{sec:concl}, we offer our conclusions
and outlook for future research directions. In the \ref{sec:bosonization_and_mapping},
we give more details about the technical aspects of our calculations.

\section{Model\label{sec:model}}

\begin{figure}
\begin{centering}
\includegraphics[bb=0bp 300bp 650bp 600bp,clip,scale=0.4]{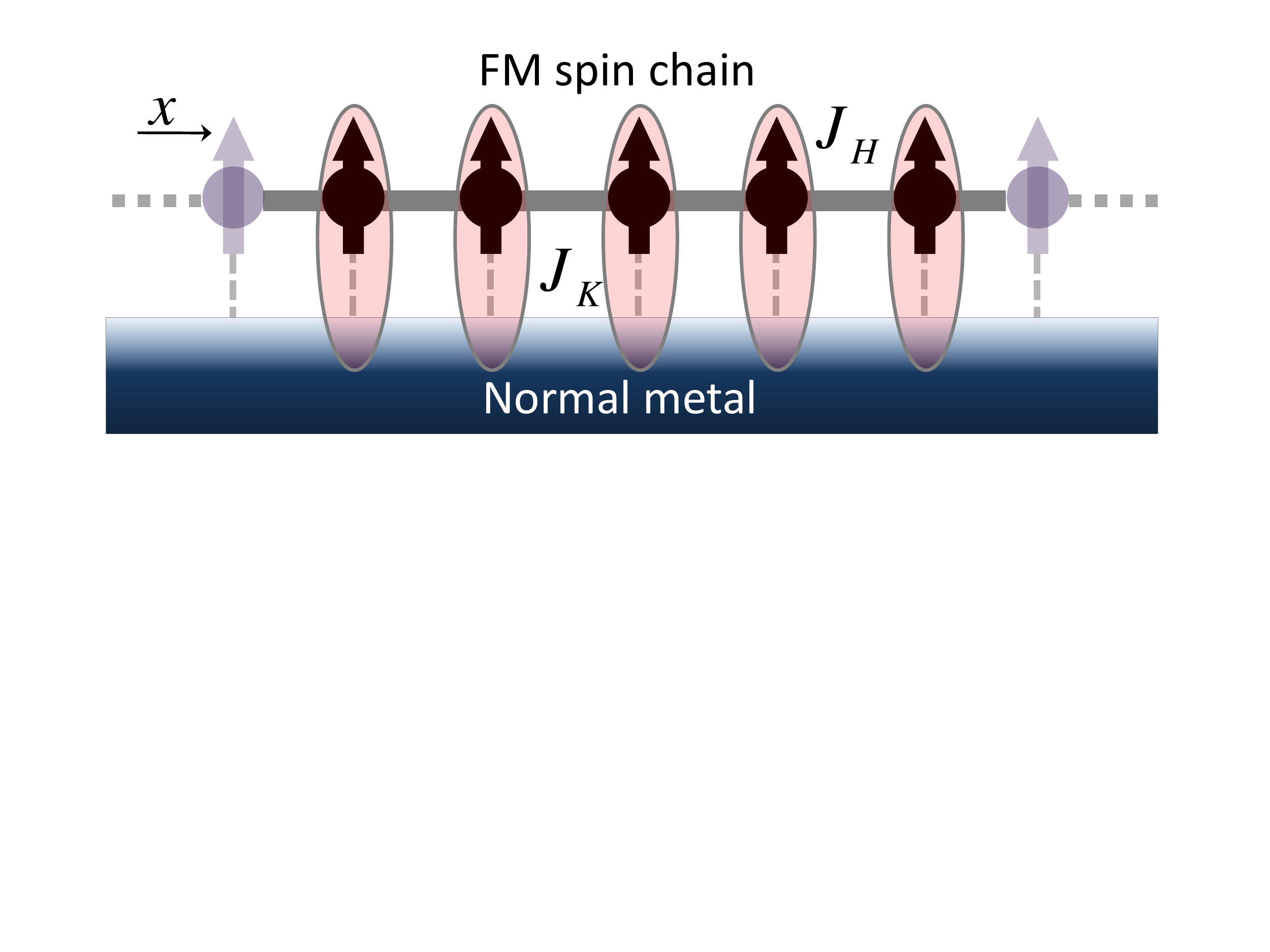} 
\par\end{centering}

\caption{\label{fig:system}Graphical representation of a FM spin chain deposited
on the top of a metallic surface. Kondo-screening of the individual
spins at low temperatures $T\ll T_{K}$ tends to destroy the FM long-range
order and induce a quantum phase transition to a paramagnetic phase
(cf. Fig. \ref{fig:mf_phase_diag}). }
\end{figure}

Let us consider a chain of magnetic atoms deposited on top of metallic
surface (see Fig. \ref{fig:system}). We model the magnetic interactions
between the atoms are using the spin-$\frac{1}{2}$ \textit{XXZ} Hamiltonian:
\begin{equation}
H_{0}  = - \sum_{i=1}^{N} \left[ J_{H}^{\perp}\left(S_{i}^{x}S_{i+1}^{x}+S_{i+1}^{y}S_{i}^{y}\right)+J_{H}^{z}S_{i}^{z}S_{i+1}^{z} 
\right].\label{eq:h0}
\end{equation}
 In the above expression, we have assumed anisotropic exchange interactions,
which originate from the reduced symmetry of the surface crystal environment,
as well as from spin-orbit interactions of the electrons in the metallic
substrate~\cite{Carbone11_Self-assembled_magnetic_networks_on_surfaces}.
Assuming that the impurity spin is $S=\frac{1}{2}$ implies that the
single-ion anisotropy $D\sum_{i}(S_{i}^{z})^{2}$ reduces to a constant
term. Furthermore, for $S=\frac{1}{2}$, exchange terms like $\sum_{i}\left(\mathbf{S}_{i}\cdot\mathbf{S}_{i+1}\right)^{2}$
can be expressed in terms of $\sum_{i}\left(\mathbf{S}_{i}\cdot\mathbf{S}_{i+1}\right)$.
However, for impurities with $S>\frac{1}{2}$ such terms must be taken
into account, but we shall not pursue their analysis here. 

The nature of the ground state of Eq.~(\ref{eq:h0}) depends crucially
on the ratio $\Delta=J_{H}^{z}/J_{H}^{\perp}$, where we assume $J_{H}^{\perp}\geqslant0$ and $J^z_H > 0$.
For $\Delta > 1$ $\left(\Delta < -1\right)$, Eq. (\ref{eq:h0}) describes
an easy-axis (i.e. Ising-like) ferromagnet (anti-ferromagnet). In
this system, the spin waves (magnons) and their bound states are separated
from the ground state by a gap~\cite{Schneider82_spectrum_FM_XXZ,Mattis_The_theory_of_magnetism_I}.
On the other hand, for $\left|\Delta\right|<1$, the spin chain Hamiltonian
(\ref{eq:h0}) exhibits a \textit{XY} phase with a gapless spin-wave
spectrum~\cite{affleck_houches,giamarchi_book_1d}. However, as anticipated
in the introduction, in what follows we will focus only on the FM
Ising limit $\Delta > 1$ of (\ref{eq:h0}), and we refer the reader
to Ref. \cite{Lobos12_Dissipative_XY_chain} for a complementary study
in the case of easy-plane anisotropy, where the spin chain exhibits
an \textit{XY} phase. Nevertheless, we note that the results reported
in Section~\ref{sec:mft} are also applicable to the antiferromagnetic
(AFM) regime corresponding to $\Delta < -1$ (see Section~\ref{sec:concl}).
Finally, for the sake of simplicity, we have neglected longer-range
magnetic exchange interactions in Eq. (\ref{eq:h0}), may be present
due to the Rudermann-Kittel-Kasuya-Yosida (RKKY) mechanism mediated
by the metallic substrate \cite{ruderman54}. We note that these interactions
can be easily included, as long as they do not modify the nature of
the Ising FM ground state.

The coupling between the spin chain and the metallic substrate is
described by an anisotropic Kondo-exchange Hamiltonian \cite{hewson}:
\begin{eqnarray}
H_{K} & = & \sum_{i=1}^{N}J_{K}^{\perp}\left[S_{i}^{x}s^{x}\left(\mathbf{R}_{i}\right)+S_{i}^{y}s^{y}\left(\mathbf{R}_{i}\right)\right]+J_{K}^{z}S_{i}^{z}s^{z}\left(\mathbf{R}_{i}\right),\label{eq:hk}
\end{eqnarray}
 where $s^{a}\left(\mathbf{R}_{i}\right)$ ($a=x,y,z$) is the spin-density
of conduction electrons at the position of the $i$-th spin impurity
in the surface $\mathbf{R}_{i}=i.a_{0}$ (with $a_{0}$ the lattice
parameter of the chain). Note that this model bears some similarity
with the Kondo-lattice model, which is used to describe the properties
of heavy-fermion systems \cite{hewson,Lohneysen99_Ce_review,Si10_Heavy_fermions_and_phase_transitions}.
In the case of impurities with $S>\frac{1}{2}$, the Kondo exchange
involves more channels (with channel-dependent couplings) and results
in a dynamical screening of the impurity spins occurring in several
stages, involving different Kondo temperatures~\cite{nozieres_multichannel}%
. We shall not pursue such an analysis here.

\begin{figure}
\begin{centering}
\includegraphics[bb=0bp 200bp 650bp 600bp,clip,scale=0.4]{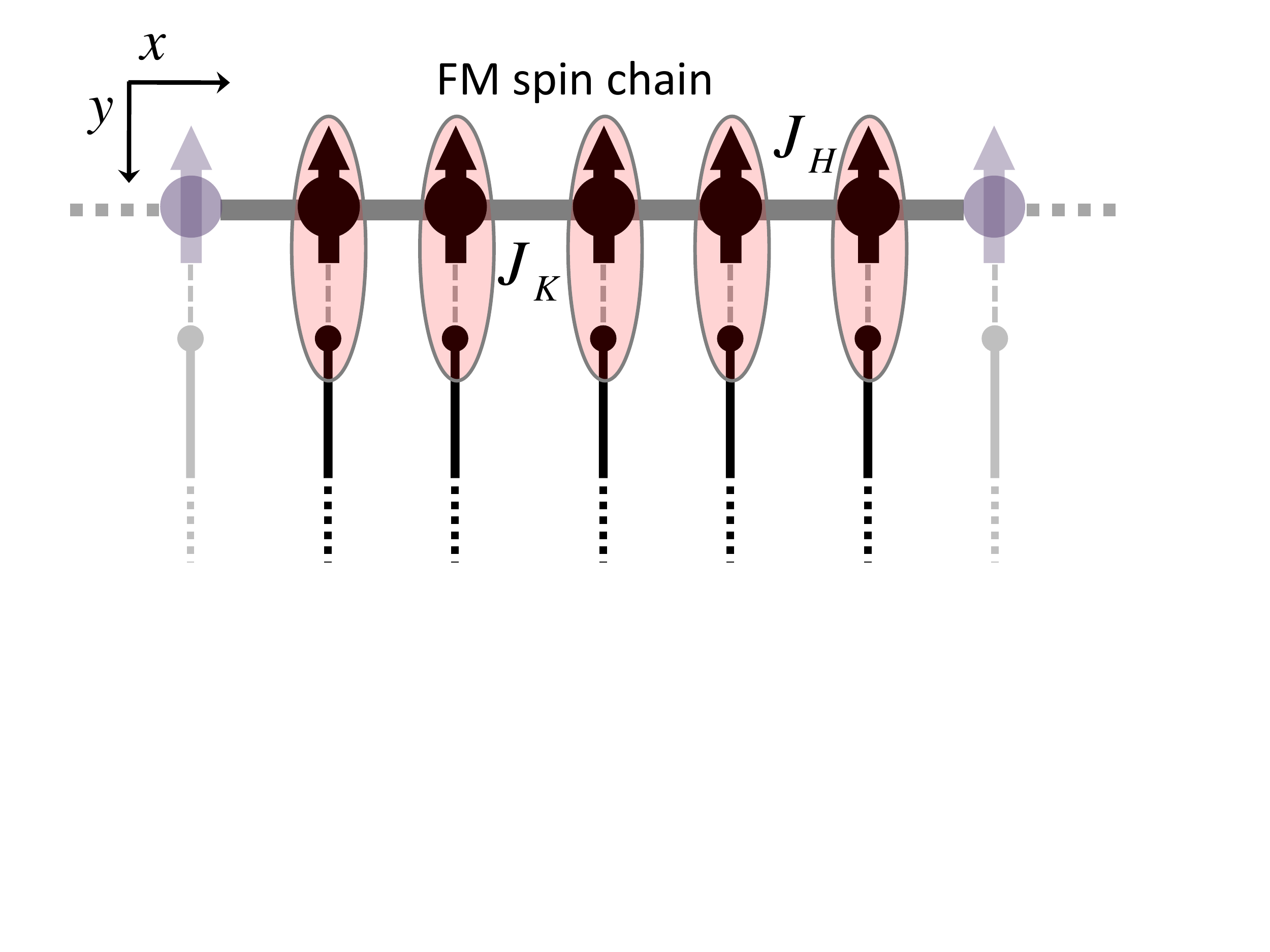} 
\par\end{centering}

\caption{\label{fig:local-bath-approximation} Local bath approximation. For
impurities embedded in a 2D or 3D metallic substrate, the interference
of Kondo screening clouds is negligible when the impurities are separated
by a few Fermi wave numbers $k_{F}^{-1}$ \cite{Andreani93_Two-impurity_Anderson_model_variational,Barzykin00_Kondo_cloud,Simonin07_Kondo_cloud},
and therefore can be considered as independently screened \cite{Lobos12_Dissipative_XY_chain}.}
\end{figure}

Despite its simplicity, the model introduced above still represents
a formidable theoretical challenge, and in order to make analytical
progress we shall introduce some simplifying assumptions. We approximate
the metallic substrate by a collection of one-dimensional metals (cf.
Fig.~\ref{fig:local-bath-approximation}) described by the Hamiltonian
\begin{eqnarray}
H_{F} & = & \sum_{i=1}^{N}\sum_{k,\alpha}\left(\epsilon_{k}-\mu\right)\: c_{\alpha i}^{\dagger}\left(k\right)c_{\alpha i}\left(k\right),\label{eq:hf}
\end{eqnarray}
 where $c_{\alpha i}^{\dagger}\left(k\right)$ creates an electron
with linear momentum $k$ and spin projection $\alpha$. The index
$i$ indicates that the electron reservoir described by the operators
$c_{\alpha i}^{\dag}\left(k\right),c_{\alpha i}\left(k\right)$ is
only coupled to the impurity located at $\mathbf{R}_{i}$ and not
to others with $j\neq i$. For this reason, we dubbed this approximation
the ``local bath approximation'' (LBA)~\cite{Lobos12_Dissipative_XY_chain}.
Within the LBA, the spin-density operator reads: 
\begin{eqnarray}
s^{a}\left(\mathbf{R}_{i}\right) & = & \sum_{\alpha,\beta}c_{\alpha i}^{\dagger}\left(k\right)\left(\frac{\sigma^{a}}{2}\right)_{\alpha\beta}c_{\beta i}\left(k\right),
\end{eqnarray}
 where $\sigma^{a}$ are the Pauli matrices.

The dimensionality of the substrate is crucial for the justification
of the LBA. When the magnetic impurities embedded in a high-dimensional
(2D or 3D) metallic substrate are separated by a distance of the order
of a few $k_{F}^{-1}$, with $k_{F}$ the Fermi wavevector of conduction
electrons, the interference of Kondo screening clouds that dynamically
quench the magnetic moments, becomes negligible \cite{Andreani93_Two-impurity_Anderson_model_variational,Barzykin00_Kondo_cloud,Simonin07_Kondo_cloud}.
Consequently, at distances $a_{0}\gtrsim k_{F}^{-1}$ the impurities
can be considered as independently screened \cite{Lobos12_Dissipative_XY_chain}.
This is an important difference with respect to the case of a strictly
1D metallic substrate, as in the case of the 1D Kondo-lattice model
\cite{zachar_kondo_chain_toulouse}. In that case, the single Kondo-impurity
limit is reached only at distances $a_{0}\gg\xi_{K}\sim v_{F}/T_{K}$.
Since the Kondo temperature $T_{K}$ is an exponentially small scale,
in the purely 1D geometry the single-impurity regime is only reached
at extremely dilute impurity-spin concentrations \cite{Andreani93_Two-impurity_Anderson_model_variational,Barzykin00_Kondo_cloud,Simonin07_Kondo_cloud}.

In actual experiments, although the magnetic nanostructures are built
on the top of a metallic surface, $k_{F}$ is determined by the 3D
bulk electron density due to a non-vanishing overlap between the bulk
and surface conduction states \cite{knorr02}. For instance, the Fermi
wave number for Au bulk free conduction electrons is $\pi/k_{F}^{\textrm{bulk}}=2.6\textrm{\AA}$,
while for Au(111) surface states is $\pi/k_{F}^{\textrm{surf}}=16.6\textrm{\AA}$
\cite{ujsahy00}. This fact dramatically increases the range of applicability
of the LBA. The physical picture provided by the LBA is also supported
by the behavior of the STS Fano line shapes in experiments on magnetic
Co atoms deposited on Cu$(100)$ and separated by distances $a_{0}>8\:\textrm{\AA}$,
which are identical to the single-impurity STS line shapes \cite{Wahl07_Exchange_Interaction_between_Single_Magnetic_Adatoms}.

In addition, although in the actual experimental realizations the
limit $a_{0}\gg k_{F}^{-1}$ might not be strictly realized, the model
resulting from applying the LBA to Eqs.~(\ref{eq:h0},\ref{eq:hk})
can be considered as the simplest ``toy model'' that captures the
competition between the Kondo physics, which is a local quantum critical
phenomenon, and the magnetic interactions along the chain \cite{Lobos12_Dissipative_XY_chain}.
Concerning the latter, it is worth noting that the LBA cannot reproduce
the non-local RKKY exchange coupling mediated by the metal and arising
at order $\mathcal{O}\left(J_{K}^{2}\right)$. As mentioned before,
this is not a crucial drawback, since one can always redefine the
couplings $J_{H}^{\perp},J_{H}^{z}$ in Eq. (\ref{eq:h0}) to take
the RKKY contribution explicitly into account. More importantly, what
eventually justifies the separate treatment of RKKY and Kondo interactions
is that, although they both originate in the same Hamiltonian (\ref{eq:hk}),
the RKKY interaction results from electronic states deep inside the
Fermi sea and is a static coupling, while the Kondo effect is purely
dynamical effect originated predominantly in the Fermi surface~\cite{CastroNeto00_NFL_in_U_and_Ce_alloys,Si10_Heavy_fermions_and_phase_transitions}.

\section{Mean-field analysis of the $T\rightarrow0$ phase diagram\label{sec:mft}}

In order to gain insight into the low-temperature phase diagram of
the system, we now focus on the limit $\Delta > 1$ of Eq. (\ref{eq:h0}),
describing an isolated FM Ising spin chain. In that case, the ground
state of the isolated chain is separated from the single-magnon excitation
spectrum by a gap $E_{g}= J_{H}^{z} \left(1- \Delta^{-1}\right)$ and
by a gap $E_{g}^{\prime}= J_{H}^{z} \left(1-\Delta^{-2}\right)>E_{g}$
from the two-magnon bound states ~\cite{Schneider82_spectrum_FM_XXZ,Mattis_The_theory_of_magnetism_I}.
This means that the long-range FM order in the isolated chain is stable
at low temperatures and, to a first approximation, it is safe to take
$J_{H}^{\perp}=0$ in Eq. (\ref{eq:h0}).

We now introduce the coupling to the substrate Eq. (\ref{eq:hk}),
which has important consequences for the low-temperature behavior
of the the spin chain. For instance, the nature of the ferromagnetic
to paramagnetic phase transition is profoundly modified. As we show
in the \ref{sec:bosonization_and_mapping}, in the case $J_{H}^{\perp}=0$
the model for the Ising chain coupled to the substrate can be mapped
onto the 1D dissipative quantum Ising model, which is in the universality
class of the classical 3D (instead of 1D) Ising model. In the following,
we exploit this fact to introduce a mean-field (MF) approach, which
therefore provides a good description of the phases of the spin chain
in the (easy-axis) Ising limit.

We introduce the MF approximation to the Ising term in Eq. (\ref{eq:h0})
by making the replacement $S_{i}^{z}S_{i+1}^{z}\rightarrow S_{i}^{z}\left\langle S_{i+1}^{z}\right\rangle +\left\langle S_{i}^{z}\right\rangle S_{i+1}^{z}-\left\langle S_{i}^{z}\right\rangle \left\langle S_{i+1}^{z}\right\rangle $,
which leads to the MF Hamiltonian of the spin chain: 
\begin{eqnarray}
H_{0}^{MF} & = & \frac{Nh_{\text{eff}}^{2}}{4 J_{H}^{z}}-\sum_{i=1}^{N}h_{\text{eff}}S_{i}^{z},\label{eq:h0_mf}
\end{eqnarray}
 where 
\begin{eqnarray}
h_{\text{eff}} & = & 2 J_{H}^{z} \left\langle S_{i}^{z}\right\rangle \label{eq:weiss_field}
\end{eqnarray}
 is the local Weiss field. The MF Hamiltonian of the system is therefore
written as 
\begin{eqnarray}
H^{MF} & = & \sum_{i=1}^{N}\mathcal{H}_{i}^{MF}+\frac{Nh_{\text{eff}}^{2}}{4J_{H}^{z}},\label{eq:H_MF_total}\\
\mathcal{H}_{i}^{MF} & = & J_{K}^{\perp}\left[S_{i}^{x}s^{x}\left(\mathbf{R}_{i}\right)+S_{i}^{y}s^{y}\left(\mathbf{R}_{i}\right)\right]+J_{K}^{z}S_{i}^{z}s^{z}\left(\mathbf{R}_{i}\right)\nonumber \\
 &  & -h_{\text{eff}}S_{i}^{z}+\mathcal{H}_{Fi},\label{eq:H_MF_local}
\end{eqnarray}
 with $\mathcal{H}_{Fi}=\sum_{k\alpha}\epsilon_{k}\: c_{\alpha i}^{\dagger}\left(k\right)c_{\alpha i}\left(k\right)$
describing the $i$-th conduction electron bath. In the following
we drop the lattice sub-index $i$ in $\mathcal{H}_{i}^{MF}$ and
$\mathcal{H}_{Fi}$ in order to lighten the notation.

We note that the MF approximation together with the LBA leads to a
set of independent Kondo-impurity problems in an effective magnetic
field $h_{\text{eff}}$, which is a quantity that must be obtained
self-consistently. The MF thermodynamical properties of the spin-chain
can therefore be obtained from the free-energy $f\left(T,h_{\text{eff}}\right)=-T\ln\text{Tr }\left[e^{-\mathcal{H}^{MF}\left(h_{\text{eff}}\right)/T}\right]$
for a single Kondo impurity in a magnetic field $h_{\text{eff}}$.
Note that the Weiss field $h_{\mathrm{eff}}$ does not act upon the
conduction electrons, it only acts on the impurity. The mean-field
self-consistency condition is obtained from the relation between the
impurity magnetization, 
\begin{equation}
m_{\text{imp}}\left(T,h_{\text{eff}}\right)=\left\langle S^{z}\right\rangle =-\frac{\partial f\left(T,h_{\text{eff}}\right)}{\partial h_{\text{eff}}}
\end{equation}
 and the definition of the Weiss field Eq. (\ref{eq:weiss_field}):
\begin{eqnarray}
m_{\text{imp}}\left(T,h_{\text{eff}}\right) & = & \frac{h_{\text{eff}}}{2J_{H}^{z}}.\label{eq:mf_self_consistent_cond}
\end{eqnarray}

  To make further progress we now need to know the function $m_{\text{imp}}\left(T,h_{\text{eff}}\right)$
for a single Kondo impurity. Introducing the Abelian bosonization
method \cite{giamarchi_book_1d} to describe the metallic 1D chains
in Fig. \ref{fig:local-bath-approximation}, and the unitary Emery-Kivelson
transformation \cite{emery_kivelson_kondo_review} {[}cf. Eq. (\ref{eq:transf_U}){]},
the anisotropic Kondo model can be mapped onto the resonant-level
model \cite{schlottmann_transformation_kondo,wiegmann_kondo_line,tsvelick_kondo_review}.
The resonant-level model is exactly solvable when the parameter $J_{K}^{z}$
is chosen at the so-called ``Toulouse line'' \cite{toulouse_kondo_line},
defined by $\tan^{-1}\left(\pi\rho_{0}b_{0}J_{K}^{z*}/4\right)=\frac{\pi}{2}\left(1-\frac{1}{\sqrt{2}}\right)$,
where $\rho_{0}$ and $b_{0}$ are the density of states and lattice
parameter of the bosonized 1D chains, respectively (cf. \ref{sec:bosonization_and_mapping}).
In combination with the renormalization group (RG) approach \cite{anderson70},
which describes the flow of parameters $\left\{ J_{K}^{z}\left(\ell\right),J_{K}^{\perp}\left(\ell\right)\right\} $
of the anisotropic Kondo model as a function of the conduction-band
cutoff parametrized by $\Lambda\left(\ell\right)=\Lambda_{0}e^{-\ell}$,
the mapping to the resonant-level model allows to describe the physical
properties of the strong coupling limit of the Kondo model. One starts
with the ``bare'' couplings $\left\{ J_{K}^{z},J_{K}^{\perp}\right\} =\left\{ J_{K}^{z}\left(\ell=0\right),J_{K}^{\perp}\left(\ell=0\right)\right\} $,
and renormalizes up to the point where $J_{K}^{z}\left(\ell^{*}\right)=J_{K}^{z*}$,
and at that point we use the exact solution of the resonant-level
model \cite{giamarchi_book_1d}. However, we stress that $m_{\text{imp}}\left(T,h_{\text{eff}}\right)$
can be obtained in a more general case (and with more sophistication)
using any ``impurity-solver'' method, such as the numerical renormalization
group (NRG)~\cite{hewson}. The mapping to the resonant-level model
yields the result \cite{schlottmann_transformation_kondo,wiegmann_kondo_line,tsvelick_kondo_review}
\begin{eqnarray}
m_{\text{imp}}\left(T,h_{\text{eff}}\right) & = & \frac{1}{\pi}\text{Im}\:\left[\psi\left(\frac{1}{2}+\frac{T_{K}}{2\pi T}+\frac{ih_{\text{eff}}}{2\pi T}\right)\right],\label{eq:mz_resonant-level}
\end{eqnarray}
 where $\psi\left(z\right)$ is the digamma function \cite{abramowitz_math_functions}
and the broadening of the resonant level $\Gamma=\rho_{0}b_{0}\left(\frac{J_{K}^{\perp}\left(\ell^{*}\right)}{4\sqrt{\pi}}\right)^{2}$
can be interpreted as $T_{K}\sim\Gamma$.

\begin{figure}
\begin{centering}
\includegraphics[scale=1.1]{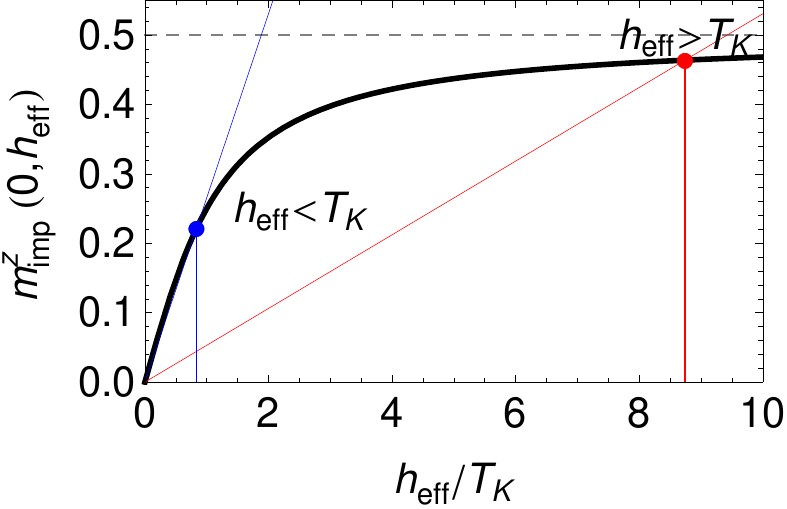} 
\par\end{centering}

\caption{\label{fig:mft}Graphical solution of the mean-field equation, Eq.~\ref{eq:mf_self_consistent_cond},
for $T=0$. For $J_{H}^{z}$ larger than the critical value there
are two kinds of mean-field solutions: those for which $h_{\text{eff}}<T_{K}$
and those for which $h_{\text{eff}}>T_{K}$. In the first case, the
coupling between the impurities does not suffice to destroy the Kondo
singlet. However, in the latter case, the coupling is strong enough
to break the Kondo singlets.}
\end{figure}

\begin{figure}
\begin{centering}
\includegraphics[scale=0.9]{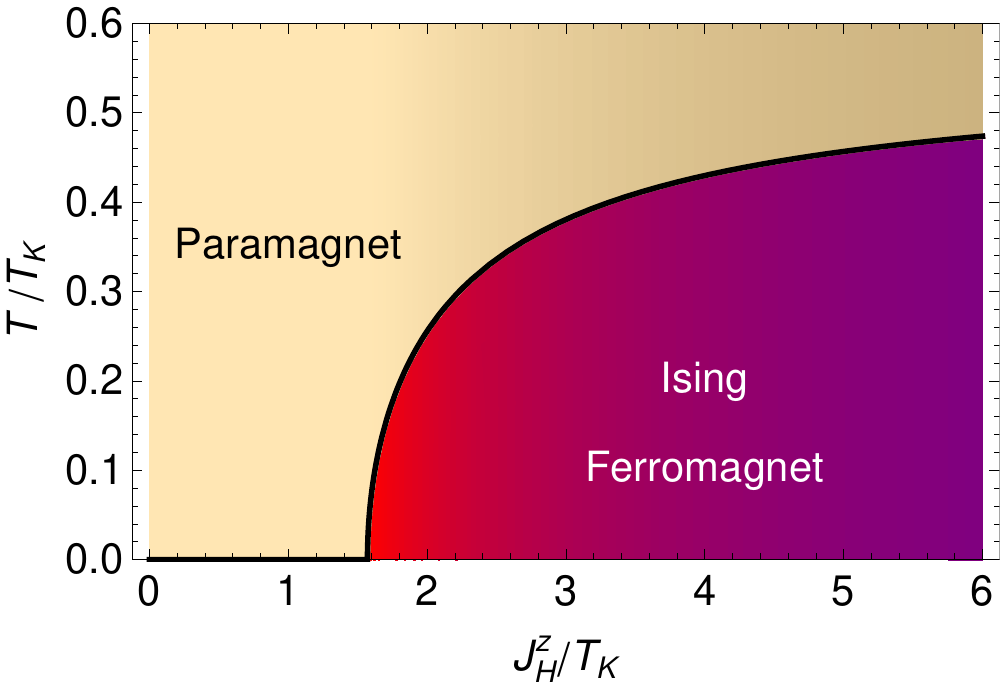} 
\par\end{centering}

\caption{\label{fig:mf_phase_diag}Mean-field phase diagram of the Ising chain
coupled to the metallic substrate. At low enough temperatures, and
in the regime $J_{H}^{z}\ll T_{K}$ quantum fluctuations induced by
the (Kondo) coupling to the substrate completely destroy the long-range
magnetic order.}
\end{figure}

The phase boundary between the FM phase where $\left\langle S^{z}\right\rangle \propto h_{\text{eff}}\neq0$,
and the paramagnetic phase with $\left\langle S^{z}\right\rangle =0$
is obtained from (\ref{eq:mf_self_consistent_cond}) in the limit
$h_{\text{eff}}\rightarrow0$ (cf. Fig~\ref{fig:mft}). For $\left\{ T/T_{K},h_{\text{eff}}/T_{K}\right\} \ll1$,
the impurity exhibits a characteristic (Pauli) paramagnetic response
$m_{\text{imp}}\left(T,h_{\text{eff}}\right)=\chi_{\text{imp}}\left(T\right)\, h_{\text{eff}}$,
where%

\begin{eqnarray}
\chi_{\text{imp}}\left(T\right) & = & \frac{1}{2\pi^{2}T}\psi^{\prime}\left(\frac{1}{2}+\frac{T_{K}}{2\pi T}\right),\label{eq:chi_T}\\
 & \approx & \chi_{\text{imp}}\left(0\right)\left[1-\frac{\pi^{2}}{3}\left(\frac{T}{T_{K}}\right)^{2}+\mathcal{O}\left(\frac{T}{T_{K}}\right)^{4}\right],\label{eq:chi_T_approx}
\end{eqnarray}
where $\chi_{\text{imp}}\left(0\right)=1/\pi T_{K}$ is the magnetic
susceptibility of the impurity at $T=0$. On the other hand, for $h_{\text{eff}}\gg\text{min}\left\{ T_{K},T\right\} $,
$m_{\mathrm{imp}}\left(T,h_{\text{eff}}\right)$ approaches the saturation
limit of $1/2$ as

\begin{eqnarray}
m_{\mathrm{imp}}\left(T,h_{\text{eff}}\right) & \simeq & \frac{1}{\pi}\arctan\left[\frac{h_{\text{eff}}}{2\pi T+T_{K}}\right].\label{eq:mz_saturation}
\end{eqnarray}
The above result can be also obtained using the scaling equations,
and it is modified by anisotropy at the Toulouse point. Thus, a non-vanishing
solution for $h_{\text{eff}}$ is found from Eq.~(\ref{eq:mf_self_consistent_cond})
in the limit $h_{\text{eff}}\rightarrow0$ 
\begin{eqnarray}
\frac{1}{2J_{H}^{z}} & < & \chi_{\text{imp}}\left(0\right)\left[1-\frac{\pi^{2}}{3}\left(\frac{T}{T_{K}}\right)^{2}\right].\label{eq:non-trivial}
\end{eqnarray}
 From the above considerations, the phase boundary is therefore given
by the critical line

\begin{eqnarray}
\frac{T_{c}}{T_{K}} & = & \frac{\sqrt{3}}{\pi}\sqrt{1-\frac{\pi T_{K}}{2 J_{H}^{z}}},\label{eq:Tc}
\end{eqnarray}
 and from here we can obtain the MF phase diagram of the spin chain
(See Fig. \ref{fig:mf_phase_diag}).

It is interesting to note that, in the FM phase, as $J_{H}^{z}/T_{K}$
increases from the phase boundary, the value of the Weiss field $h_{\mathrm{eff}}$
crosses over from a regime $h_{\mathrm{eff}}\ll T_{K}$ where the
Kondo singlet is still robust, to the regime of large $h_{\mathrm{eff}}\gg T_{K}$
where the Kondo singlet is destroyed (cf. Fig.~\ref{fig:mft}). Therefore,
the above mean-field theory yields a regime where the Kondo effect
and a FM LRO coexist. In the following Section, we shall return to
this point again.

It is also worth mentioning that the result in Eq.~(\ref{eq:Tc})
clearly contrasts with the case of the isolated Ising chain, for which
the analytical form of $T_{c}$ Eq. (\ref{eq:Tc}), is very different
from the expression for $T_{c}$ encountered in the study of the classical
Ising model, where $T_{c}\propto J_{H}^{z}$ \cite{ashcroft_mermin_book}.
Physically, in the present case, the spin-flip processes induced by
the Kondo coupling to the metal Eq. (\ref{eq:hk}) can be considered
as quantum fluctuations that tend to destroy the LRO in the Ising
chain as $T\rightarrow0$ for $J_{H}^{z}/T_{K}\lesssim1.6$ (See Fig.
\ref{fig:mf_phase_diag}).

In the following Section we characterize the excitations of the ordered
FM phase.

\section{Magnetic excitations in the FM phase\label{sec:holtsein}}

We next focus on the FM phase, in the regime $J_{H}^{z}/T_{K} \gtrsim 1$
(right-bottom region in Fig. \ref{fig:mf_phase_diag}), where the
spectrum of the system is gapped and the effective Weiss field $h_{\text{eff}}$
is large enough so as to inhibit the Kondo screening of the impurities.
We shall focus in particular on single magnon excitations that can
be created by probes (such like spin-polarized electron energy loss
spectroscopy, SPEELS, see Section \ref{sec:experiments}) that couple
weakly to the ferromagnetic chains. Such probes are assumed to produce
single spin-flips, and thus the calculation of the system response
can be carried out within linear response theory. This approximation
allows us to neglect the effect of magnon bound states in the spectrum
of the FM \textit{XXZ} chain. Thus, the low-lying excitations involving
a single spin-flip can be described using the Holstein-Primakoff (HP)
representation of the spin operators \cite{auerbach_book_spins}:
\begin{eqnarray}
S_{i}^{+} & = & b_{i}^{\dagger}\sqrt{2S-b_{i}^{\dagger}b_{i}}=\sqrt{2S}\left[b_{i}^{\dagger}+O\left(\frac{1}{S}\right)\right],\label{eq:Sp_HP}\\
S_{i}^{-} & = & \sqrt{2S-b_{i}^{\dagger}b_{i}}b_{i}=\sqrt{2S}\left[b_{i}+O\left(\frac{1}{S}\right)\right],\label{eq:Sm_HP}\\
S_{i}^{z} & = & \left(b_{i}^{\dagger}b_{i}-S\right),\label{eq:Sz_HP}
\end{eqnarray}
 where $b_{i},b_{i}^{\dagger}$ are bosonic annihilation and destruction
operators, obeying $\left[b_{i},b_{j}^{\dagger}\right]=\delta_{ij}$,
which physically represent a single spin-flip. For the \textit{XXZ}
chain introduced in Sec.~\ref{sec:model}, we should take $S=\frac{1}{2}$
in the above expressions. However, below we shall work assuming that
$S$ is large, which is known to provide an accurate description of
the lowest energy excitations~\cite{auerbach_book_spins}. Assuming
$S$ large means that our results also apply to spin chains with $S>\frac{1}{2}$,
provided the Hamiltonian is of the \textit{XXZ} form and we assume
that the Kondo-exchange coupling with the substrate describes the
strongest channel of substrate electrons that couple to each impurity
along the chain. Thus, working to leading order in $S$ (i.e. in the
large-$S$ approximation), we find: 
\begin{eqnarray}
H_{0}^{HP} & \simeq &  \sum_{i=1}^{N}\left[-J_{H}^{z}S^{2}- J_{H}^{\perp}S\left(b_{i}^{\dagger}b_{i+1}+b_{i+1}^{\dagger}b_{i}\right)\right.\nonumber \\
 &  & \left. + J_{H}^{z} S \left(b_{i}^{\dagger}b_{i}+b_{i+1}^{\dagger}b_{i+1}\right)\right],\label{eq:h0_HP}
\end{eqnarray}
 which can be easily diagonalized in a running-wave basis 
\begin{eqnarray}
b_{i} & = & \frac{1}{\sqrt{N}}\sum_{q}e^{iqR_{i}}\: b_{q},\label{eq:Fourier_transf}
\end{eqnarray}
 and leads to

\begin{eqnarray}
H_{0}^{HP} & = & \sum_{q}E_{q}b_{q}^{\dagger}b_{q}-NJ_{H}^{z}S^{2}+\mathcal{O}\left(1\right),\label{eq:h0_HP2}
\end{eqnarray}
 where 
\begin{eqnarray}
E_{q} & = & 2S J_{H}^{z} - 2S J_{H}^{\perp}\cos\left(qa_{0}\right),\nonumber \\
 & \simeq & 2S \left(J_{H}^{z}   - J_{H}^{\perp}\right) +2S  J_{H}^{\perp} \frac{q^{2}a^2_0}{2}+\mathcal{O}
 \left(q^{4}\right),\label{eq:E_q}
\end{eqnarray}
 is the dispersion relation for magnons, and $a_{0}$ is the lattice
parameter. In addition, in Eq. (\ref{eq:h0_HP2}) we have neglected
terms of $\mathcal{O}\left(1\right)$, which involve interactions
between the magnons. From Eq. (\ref{eq:E_q}), we see that for $J_{H}^{z}>J_{H}^{\perp} > 0$ (i.e. $\Delta  > 1$)
the excitation spectrum of the spin chain is gapped, and the spin
waves exhibit a quadratic dispersion for momentum $q \to 0$.

Let us next consider on the Kondo-exchange coupling to the metal,
Eq. (\ref{eq:hk}). We shall treat this coupling using the HP representation
in the large-$S$ limit:

\begin{eqnarray}
H_{K}^{HP} & \simeq & \sum_{i=1}^{N}\left[J_{K}^{\perp}\sqrt{S/2}\left(b_{i}^{\dagger}s^{-}\left(\mathbf{R}_{i}\right)+s^{+}\left(\mathbf{R}_{i}\right)b_{i}\right)\right.\nonumber \\
 &  & \left.+\left(b_{i}^{\dagger}b_{i}-S\right)s^{z}\left(\mathbf{R}_{i}\right)\right].\label{eq:hk_HP}
\end{eqnarray}
 As we will see, although this Hamiltonian no longer describe the
Kondo effect (we recall that we are working in the limit of large
$h_{\text{eff}}$), the coupling to the gapless degrees of freedom
in the metal still can have important consequences to the spin chain.
To make further progress, we now need to integrate out the electronic
degrees of freedom in the conduction band. To this end, we shall rely
upon the coherent-state functional integral representation of the
partition function \cite{negele_book}:

\begin{eqnarray}
Z & = & \int\prod_{i=1}^{N}\mathcal{D}\left[\bar{b}_{i},b_{i}\right]\; e^{-\mathcal{S}_{\mathrm{eff}}^{HP}\left[\bar{b},b\right]},
\end{eqnarray}
 where $\mathcal{S}_{\mathrm{eff}}^{HP}\left[\bar{b}_{i},b_{i}\right]$
is the effective Euclidean action of the spin chain%
\begin{eqnarray}
\mathcal{S}_{\mathrm{eff}}^{HP}\left[\bar{b},b\right] & = & \mathcal{S}_{0}^{HP}\left[\bar{b},b\right]+\mathcal{S}_{\text{diss}}^{HP}\left[\bar{b},b\right],\label{eq:S_eff_definition}
\end{eqnarray}
{} where the first term is the Euclidean action corresponding to (\ref{eq:h0_HP2})
\begin{eqnarray}
\mathcal{S}_{0}^{HP}\left[\bar{b},b\right] & = & \sum_{q}\int_{0}^{\hbar\beta}d\tau\bar{b}_{q}\left(\tau\right)\left[\partial_{\tau}+E_{q}/\hbar\right]b_{q}\left(\tau\right),\\
 & = & \frac{1}{\hbar^{2}\beta}\sum_{q,\omega_{n}}\left[-i\hbar\omega_{n}+E_{q}\right]\bar{b}_{q}\left(\omega_{n}\right)b_{q}\left(\omega_{n}\right),
\end{eqnarray}
 and $\mathcal{S}_{\text{diss}}^{HP}\left[\bar{b},b\right]=-\ln\left\langle \exp\left[-\int_{0}^{\hbar\beta}d\tau\; H_{K}^{HP}\left(\tau\right)\right]\right\rangle _{Fi}$
is the contribution arising from Hamiltonian (\ref{eq:hk_HP}), where
we integrate over the fermions $\left(c_{\alpha i},\bar{c}_{\alpha i}\right)$.
Note that this can be done exactly, as $H_{F}$ is quadratic (we assume
that Fermi liquid theory applies) and leads to 
\begin{eqnarray}
\mathcal{S}_{\text{diss}}^{HP}\left[\bar{b},b\right] & = & \frac{S}{4\hbar}\sum_{i=1}^{N}\int_{0}^{\hbar\beta}d\tau d\tau^{\prime}\,\left\{ \left(J_{K}^{\perp}\right)^{2}\left[\bar{b}_{i}\left(\tau\right)\chi_{F}^{\pm}\left(\tau-\tau^{\prime}\right)b_{i}\left(\tau^{\prime}\right)\right]\right.\nonumber \\
 &  & +\left(J_{K}^{z}\right)^{2}\left[\bar{b}_{i}\left(\tau\right)b_{i}\left(\tau\right)-S\right]\chi_{F}^{zz}\left(\tau-\tau^{\prime}\right)\nonumber \\
 &  & \left.\times\left[\bar{b}_{i}\left(\tau^{\prime}\right)b\left(\tau^{\prime}\right)-S\right]\right\} ,\label{eq:S_HP_diss}
\end{eqnarray}
 where we have introduced the spin correlation functions in the conduction
band:

\begin{eqnarray}
\chi_{F}^{\pm}\left(\tau\right) & = & -\frac{1}{\hbar}\langle s^{+}\left(\mathbf{R}_{i},\tau\right)s^{-}\left(\mathbf{R}_{i},0\right)\rangle_{F},\label{eq:chi1}\\
\chi_{F}^{zz}\left(\tau\right) & = & -\frac{1}{\hbar}\langle s^{z}\left(\mathbf{R}_{i},\tau\right)s^{z}\left(\mathbf{R}_{i},0\right)\rangle_{F}.\label{eq:chi2}
\end{eqnarray}
 Note that these correlation functions do not depend on $\mathbf{R}_{i}$
because all the local fermion baths are identical. Now, since these
baths are assumed to be Fermi liquids, we have that at $T=0$ and
for $\tau\gg\frac{\hbar}{\epsilon_{F}}$ (where $\epsilon_{F}$ is
the Fermi energy of the metallic substrate): 
\begin{equation}
\chi_{F}^{\pm}\left(\tau\right)\sim - \frac{\lambda^{\pm}}{\tau^{2}},\quad\chi_{F}^{zz}\left(\tau\right)\sim - \frac{\lambda^{zz}}{\tau^{2}},\label{eq:chi_long_time}
\end{equation}
 which follows from the existence of a $\sim\omega$ (where $\omega$
is the excitation energy) spectrum of particle-hole excitations in
the Fermi sea at low energies. Note that since $\chi_{F}^{\pm}$ and
$\chi_{F}^{zz}$ have units of $\left(E\times t\right)^{-1}$ (where $E$
and $t$ denote energy and time units, respectively), which stem from
the prefactor of $\hbar$ in Eq.~\ref{eq:chi1} and \ref{eq:chi2};
then $\lambda^{\pm},\lambda^{zz}$ have units $t/E$ . In Fourier
representation, we express (\ref{eq:chi_long_time}) as: 
\begin{eqnarray}
\chi_{F}^{\pm}\left(\omega_{n}\right) & \sim & \lambda^{\pm}\left|\omega_{n}\right|,\\
\chi_{F}^{zz}\left(\omega_{n}\right) & \sim & \lambda^{zz}\left|\omega_{n}\right|,
\end{eqnarray}
where $\omega_{n}=2\pi n/\beta$ are the Matsubara frequencies, and
therefore Eq. (\ref{eq:S_HP_diss}) becomes

\begin{eqnarray}
S_{\text{diss}}^{HP}\left[\bar{b},b\right] & \simeq & \frac{\left(J_{K}^{\perp}\right)^{2}S\lambda^{\pm}}{4\hbar^{2}\beta}\sum_{\omega_{n}\neq0}\sum_{i=1}^{N}|\omega_{n}|\bar{b}_{i}\left(\omega_{n}\right)b_{i}\left(\omega_{n}\right),
\end{eqnarray}
 where we have dropped the last term in Eq.~(\ref{eq:S_HP_diss})
because it amounts to a (retarded) magnon-magnon interaction. This
is consistent with neglecting magnon interactions in (\ref{eq:E_q}).
The above contribution is quadratic, and replacing it into (\ref{eq:S_eff_definition})
leads to the effective action

\begin{eqnarray}
S_{\text{eff}}^{HP}\left[\bar{b},b\right] & = &- \frac{1}{\hbar^{2}\beta}\sum_{q,\omega_{n}}\left[g_{q}\left(\omega_{n}\right)\right]^{-1}\bar{b}_{q}\left(\omega_{n}\right)b_{q}\left(\omega_{n}\right),\label{eq:S_eff_HP_final}
\end{eqnarray}
 where we have defined the magnon propagator \cite{negele_book} 
\begin{eqnarray}
g_{q}\left(\omega_{n}\right) & = & \frac{1}{i\hbar\omega_{n}-E_{q}-\alpha|\omega_{n}|},\label{eq:propagator_Matsubara}
\end{eqnarray}
 with $\alpha\propto\lambda^{\pm}\left(J_{K}^{\perp}\right)^{2}$
a dimensionless parameter quantifying the dissipation caused by the
conduction electrons on the spin chain. We next perform the analytic
continuation to real frequencies, in order to obtain the retarded
magnon propagator. Thus, replacing $i\omega_{n}\to\omega+i0^{+}$
and $|\omega_{n}|\to-i\omega$, yields 
\begin{equation}
g_{q}^{R}\left(\omega\right)=\frac{1}{\hbar\omega-E_{q}+i\alpha\omega},
\end{equation}
 Thus, the pole of the propagator becomes $\hbar\omega_{q}=E_{q}\left(1+i\alpha/\hbar\right)^{-1}\simeq E_{q}-i\frac{\alpha E_{q}}{\hbar}$,
for small $\alpha$, which implies that the coupling to the metal
Eq. (\ref{eq:hk_HP}) induces a finite lifetime $\Gamma=\hbar\alpha^{-1}$
and damping on the magnon excitations. Physically, this happens due
to the presence of gapless particle-hole excitations in the excitation
spectrum of the metallic substrate, and it can be interpreted as Landau
damping \cite{Costa03_SPEELS_theory_in_Fe_multilayers,Costa04_SPEELS_theory_for_Co_monolayer_on_Cu111,Buczek11_Landau_damping_in_ultrathin_magnets}.
Interestingly, this damping does not vanish in the limit $q\to0$.
At first sight, this may seem surprising, since in the $q\to0$ limit,
magnon excitations correspond to a uniform magnetization along the
$z$ direction. Thus, the total projection of the spin along $z$,
i.e., $S_{T}^{z}=\sum_{i=1}^{N}S_{i}^{z}$, is a conserved quantity
in the absence of coupling to the substrate. In other words, the number
of magnons $\sum_{q}b_{q}^{\dag}b_{q}$ is a conserved quantity of
Hamiltonian in (\ref{eq:h0_HP}). However, in the presence of $H_{K}^{HP}$
{[}Eq. (\ref{eq:hk_HP}){]}, $S_{T}^{z}$ is no longer conserved (i.e.
the magnons $b_{i}$ can ``leak out'' from the chain), and therefore
total magnetization fluctuations also become damped out by the coupling
to the substrate.

The above treatment is valid provided the coupling to the substrate
is weak and can be treated perturbatively. At stronger coupling, we
have to worry about the possibility that the magnetic moments of the
impurities are Kondo-screened by the substrate electrons. To deal
with this situation, we shall rely on the non-perturbative treatment
discussed in the \ref{sec:bosonization_and_mapping}. As shown there,
in the strong-coupling limit where Kondo correlations are important,
it is possible to map the model introduced in Section~\ref{sec:model}
to a dissipative quantum Ising model. When coarse-grained over distances $\gg a_{0}$,
this model can be described by a dissipative $\phi^{4}$ field theory,
whose partition function is $Z=\int\left[d\phi\right]\, e^{-\mathcal{S}\left[\phi\right]}$,
where the action $\mathcal{S}\left[\phi\right]$ is given by the expression
\cite{Pankov04_NFL_behavior_and_2D_AFM_fluctuations}: 
\begin{equation}
\mathcal{S}\left[\phi\right]=\frac{1}{2}\sum_{q,\omega_{n}}\left[\left|\omega_{n}\right|+Dq^{2}+r\right]\left|\phi\left(q,\omega_{n}\right)\right|^{2}+\frac{g}{4}\int dxd\tau\,\phi^{4}\left(x,\tau\right),\label{eq:phi4}
\end{equation}
 being $\phi\left(x,\tau\right)=\left(L\hbar\beta\right)^{-1/2}\sum_{q,\omega_{n}}\phi\left(q,\omega_{n}\right)\: e^{i\left(qx-\omega_{n}\tau\right)}$
($\beta=1/T$) is the coarse-grained magnetization along the easy
axis, i.e. $\left\langle \phi\left(x=x_{i}\right)\right\rangle =R^{-1}\sum_{j=-R/2}^{R/2}\left\langle S_{i+j}^{z}\right\rangle $,
where $1\ll R\ll N$. This result suggests that the quantum critical
point (QPC) separating the paramagnetic and ferromagnetic ground states
(cf. Fig~\ref{fig:mf_phase_diag}) can be described by this field
theory, and therefore the critical exponents around the QCP will correspond
to those of the Wilson-Fisher fixed-point of the renormalization group~(see
e.g. Ref.~\cite{sachdev_book}, for a survey). However, in deriving
the map to the dissipative quantum Ising model, we have neglected the Heisenberg
term $\sim\frac{J_{H}^{\perp}}{2}\sum_{i}\left[S_{i}^{+}S_{i+1}^{-}+\mathrm{H.c.}\right]$
from Eq.~(\ref{eq:h0}). If this term is included, we cannot rule
out that the actual values of the critical exponents near the QCP
turn out to be different from those of the Wilson-Fisher universality
class. Addressing this question in detail require a numerical investigation
of the full model, which is beyond the scope of the present work.

Nevertheless, the field theory of Eq.~(\ref{eq:phi4}), is capable
of describing the phases of the system. Focusing on the ferromagnetic
phase where $\left\langle \phi\left(x,\tau\right)\right\rangle \neq0$,
which corresponds to $r<0$, it can be seen that the magnon excitations
are indeed described by the propagator: 
\begin{equation}
\left\langle \phi^{*}\left(q,\omega_{n}\right)\phi\left(q,\omega_{n}\right)\right\rangle =\frac{1}{\left|\omega_{n}\right|+Dq^{2}+E_{g}},
\end{equation}
 where $E_{g}=2\left|r\right|\sim J_{H}^{z}-J_{H}^{\perp}$
is the energy gap. This magnon propagator indicates that the magnetic
excitations of the chain in the strong coupling regime are overdamped
by their coupling to the metallic substrate. Physically, this makes
sense as the impurity spins are screened by the metal electrons and
thus their magnetic moments undergo strong damping by the collective
nature of their screening clouds. This regime must be contrasted with
the weak-coupling regime, where we found the damping to be weak. Thus,
a physical picture emerges that allows us to distinguish the two mean-field
solutions that we briefly described at the end of Section~\ref{sec:mft}
(cf. Fig.~\ref{fig:mft}): In the weak coupling regime, the strong
Weiss field (i.e. $h_{\mathrm{eff}}\gg T_{K}$) due to the FM coupling
of the impurities along the chain does not allow for the Kondo screening
of the local moments of the atoms in the chain. As a consequence,
the magnetic excitations of the chain are only weakly affected by
their coupling to the substrate, and thus acquire a small line-width
. On the other hand, in the strong coupling regime, the Weiss field
is much weaker (i.e. $h_{\mathrm{eff}}\ll T_{K}$) and the magnetic
moments are effectively screened by the metal electrons, leading to
overdamped magnetic excitations. In the next Section, we shall study
some experimental consequences of our findings.

\begin{figure}
\begin{centering}
\includegraphics[scale=0.8]{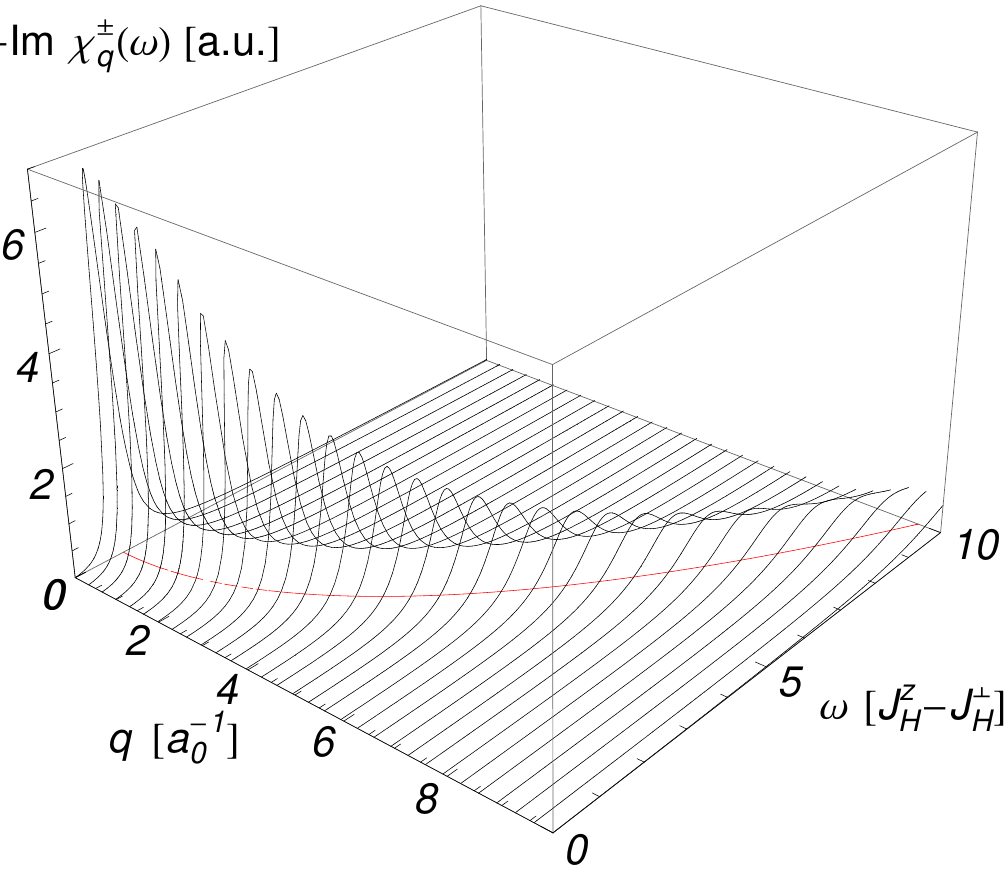} 
\par\end{centering}

\caption{\label{fig:spin_response}Spin-response $\text{Im }\chi_{q}^{\pm}\left(\omega\right)$
as a function of $q$ and $\omega$. This quantity is proportional
to the SPEELS $d^{2}P/d\Omega d\omega$ signal and provides information
about the magnon dispersion relation (red line in the bottom $q-\omega$
plane) and the FM Ising gap. The broadening of the magnon excitation
is originated in the (Kondo) coupling to the metallic substrate Eq.
(\ref{eq:hk}). }
\end{figure}

\section{Relation to SPEELS experiments\label{sec:experiments}}

In this Section, we focus on the effects of the dissipative environment
on observable quantities. The magnetic properties of low dimensional
spin systems deposited on metals have been studied with a variety
of experimental techniques, like X-ray magnetic circular dichroism
(XMCD), used to measure directly the magnetization of FM 1D spin chains
\cite{Gambardella02_Ferromagnetism_in_1D_quantum_spin_chains}, or
local probes, like spin polarized STM, which can provide information
on the magnon dispersion relation \cite{Balashov06_Magnon_excitation_with_STM}.
Here we focus on the spin-polarized electron energy-loss spectroscopy
(SPEELS) experiment \cite{Gokhale92_SPEELS_theory,Prokov09_Magnons_in_FM_monolayer_SPEELS,Zhang11_Magnon_excitations_with_SPEELS},
which provides direct information of the magnon dispersion relation.

The SPEELS cross section $d^{2}P/d\Omega d\omega$ corresponds to
the fractional number of electrons which emerge onto the solid angle
$d\Omega$ after being scattered by a magnetic excitation $\omega\left(\mathbf{q}\right)$
in the energy range $d\omega$. According to the theory of the SPEELS
experiment \cite{Gokhale92_SPEELS_theory}, $d^{2}P/d\Omega d\omega$
is related to the spin response function by

\begin{eqnarray}
\frac{d^{2}P}{d\Omega d\omega} & \propto & -\text{Im }\chi_{\mathbf{q}}^{\pm}\left(\omega\right),\label{eq:cross_section}
\end{eqnarray}
 where $\chi_{\mathbf{q}}^{\pm}\left(\omega\right)$ is the retarded
spin correlation function obtained by analytic continuation from $\chi_{\mathbf{q}}^{\pm}\left(\omega\right)=\left[\chi_{\mathbf{q}}^{\pm}\left(\omega_{n}\right)\right]_{\omega_{n}\rightarrow\omega+i0^{+}}$with
\begin{eqnarray}
\chi_{\mathbf{q}}^{\pm}\left(\omega_{n}\right) & = & \int d^{3}\mathbf{R}\int d\tau\; e^{i\mathbf{q.R}}e^{-i\omega_{n}\tau}\left\langle S^{-}\left(\mathbf{R},\tau\right)S^{+}\left(0,0\right)\right\rangle .\label{eq:chi_Fourier}
\end{eqnarray}
Using the HP representation Eqs. (\ref{eq:Sp_HP})-(\ref{eq:Sz_HP}),
we can express 
\begin{eqnarray}
\chi_{q}^{\pm}\left(\omega_{n}\right) & \approx & 2S\left\langle b_{q}\left(\omega_{n}\right)\bar{b}_{q}\left(\omega_{n}\right)\right\rangle ,\\
 & = & \frac{2S}{i\hbar\omega_{n}-\alpha|\omega_{n}|-E_{q}},
\end{eqnarray}
where we have used Eqs. (\ref{eq:S_eff_HP_final}) and (\ref{eq:propagator_Matsubara}),
and the fact that in the 1D geometry $\mathbf{q}\rightarrow q$. Introducing
this result into Eq.~(\ref{eq:cross_section}) and performing the
analytic continuation to real frequencies, we obtain:

\begin{equation}
\text{Im }\chi_{q}^{\pm}\left(\omega\right)=-\frac{\alpha\omega}{\left(\hbar\omega-E_{q}\right)^{2}+\alpha^{2}\omega^{2}}.
\end{equation}
In Fig. (\ref{fig:spin_response}) we show the spin response function
$-\text{Im }\chi_{q}^{\pm}\left(\omega\right)$ as a function of $q$
and $\omega$. The curves $-\text{Im }\chi_{q}^{\pm}\left(\omega\right)$
show resonances centered at the magnon frequencies $\omega=E_{q}$
(red line in the bottom $q-\omega$ plane), which are broadened by
the effect of $\alpha$, the dimensionless coupling to the metal.

\section{Conclusions and Outlook\label{sec:concl}}

We have studied the phase diagram and the excitation spectrum of a
magnetic chain of atoms or molecules with easy-axis ferromagnetic
interactions. We focused on a simple model of $S=\frac{1}{2}$ magnetic
impurities displaying a Kondo-exchange interaction with the substrate.
In the Ising limit, which provides a good approximation to the ground
state of an easy-axis ferromagnet, we obtained the phase diagram using
a mean-field theory approach. We find that this system exhibits two
possible phases at zero temperature: a paramagnetic phase where the
impurity spins are screened by the substrate, and a gapped ferromagnetic
phase whose excitations are damped by the magnetic (Kondo) exchange
with the metallic substrate. Using a bosonic representation for the
spins as well as various types of mathematical mappings between models,
we have also investigated the excitation spectrum in the ferromagnetic
phase, which may be accessible through probes like spin-polarized
electron energy loss (SPEELS). Although we have focused mainly on
an easy-axis ferromagnetic (FM) chain, it is worth describing how
our results can be modified in the case of the anti-ferromagnetic
(AFM) chain corresponding to the parameter regime $\Delta < -1$ of the
\textit{XXZ} chain described by Eq.~(\ref{eq:h0}). An experimental
motivation to extend our results to this regime can be found, for
instance, in Ref. \cite{DiLullo12_Molecular_Kondo_chain}. Relying
on the local bath approximation (LBA) and in the Ising limit, it is
possible to apply the transformation $S_{i}^{z}\to\left(-1\right)^{i}S_{i}^{z}$,
which maps the AFM to onto FM Ising chain. In presence of the Kondo
exchange, this transformation must be supplemented by another rotation
which takes $s_{c}^{z}\left(\mathbf{R}_{i}\right)\to\left(-1\right)^{i}s_{c}^{z}\left(\mathbf{R}_{i}\right)$
and $s^{x}\left(\mathbf{R}_{i}\right)\leftrightarrow s^{y}\left(\mathbf{R}_{i}\right)$
provided that $(-1)^{i}=-1$. For $\Delta < -1$, the excitations are
separated from the ground state by a gap $\sim\sqrt{\left(J_{H}^{z}\right)^{2}-\left(J_{H}^{\perp}\right)^{2}}$.
Using the Holstein-Primakoff representation~\cite{auerbach_book_spins},
we can compute the spectrum of an antiferromagnetic \textit{XXZ} chain,
which yields $E_{q}^{\text{AFM}}=2S\sqrt{\left(J_{H}^{z}\right)^{2}-\left(J_{H}^{\perp}\cos qa_{0}\right)^{2}}$.
The results of Sections~\ref{sec:holtsein} and \ref{sec:experiments}
carry over to the AFM ($\Delta<-1$) case provided we replace $E_{q}$
by $E_{q}^{\text{AFM}}$. A feature of the AFM regime that is also
worth noticing is that bound states of magnons are absent from the
spectrum \cite{Schneider82_spectrum_FM_XXZ,Mattis_The_theory_of_magnetism_I}.

Finally, let us discuss our outlook for future research directions.
One such direction is the study of the phases of chains of impurities
with higher (i.e. $S>\frac{1}{2}$) spin. In this case, it is also
possible to apply the same mean-field theory employed in Section~\ref{sec:mft}.
However, in order to obtain the impurity magnetization $m_{\text{imp}}\left(h_{\text{eff}},T\right)=\langle S^{z}\rangle$
we must in general resort to an impurity solver such as Wilson's numerical
renormalization group~\cite{Bulla08_NRG_review,hewson}. This is
specially true for $S>\frac{1}{2}$, which will will undergo a several
stage Kondo effect~\cite{hewson,nozieres_multichannel} and possibly
underscreening~\cite{Mehta05_Underscreened_Kondo}. Another interesting
direction will be to asses the accuracy of the local bath approximation
employed throughout. This remains quite a challenge, which will most
likely require to rely on numerical methods, such as quantum Monte
Carlo. Indeed, it is possible that this approximation only provides
an rough picture of the paramagnetic phase, which must be corrected
by the inclusion of the coupling between the different baths that
screen the magnetic impurities individually. However, we believe that
such effects will be less important in the ferromagnetic phase (specially
for large $J_{H}^{z}$), due to the existence of a gap that separates
the ferromagnetic ground state from the excitations and therefore
protects the ground state from perturbations.

\appendix

\section{Bosonization of the fermionic chains and mapping to the dissipative
quantum Ising model\label{sec:bosonization_and_mapping}}

In this Appendix, we implement the Abelian bosonization approach to
the semi-infinite fermionic 1D chains along the $y$ direction in
Fig. (\ref{fig:local-bath-approximation}) \cite{giamarchi_book_1d,gogolin_book}.
This procedure is standard and has been successfully applied to describe
the low-energy properties of the single Kondo-impurity. We refer the
reader to the standard bibliography \cite{giamarchi_book_1d,gogolin_book,schlottmann_transformation_kondo,emery_kivelson_kondo_review,Guinea85_Bosonization_of_a_two_level_system,Kotliar96_Toulouse_points_in_the_generalized_Anderson_model}.
In the present case, the LBA greatly simplifies the complexity of
the problem and enables a straightforward generalization to the case
of many-impurities, each one coupled to an independent fermionic bath.
At low temperature, in the bosonic representation \cite{Kotliar96_Toulouse_points_in_the_generalized_Anderson_model}
the Hamiltonians (\ref{eq:hf}) and (\ref{eq:hk}) become %
{} 
\begin{eqnarray}
H_{F} & = & \sum_{i,\nu=\left\{ c,s\right\} }\frac{v_{F}}{4\pi}\int_{-\infty}^{\infty}dy\;\left(\nabla\phi_{i,\nu}^{R}\left(y\right)\right)^{2},\label{eq:hf_bosonized}\\
H_{K} & = & \sum_{i}-\frac{2\delta_{s}}{\pi\rho_{0}}S_{i}^{z}\frac{\nabla\phi_{i,s}^{R}\left(0\right)}{\sqrt{2}\pi}\nonumber \\
 &  & +\frac{J_{K}^{\perp}b_{0}}{2}\left[S_{i}^{+}\frac{e^{-i\sqrt{2}\phi_{i,s}^{R}\left(0\right)}}{2\pi b_{0}}+S_{i}^{-}\frac{e^{i\sqrt{2}\phi_{i,s}^{R}\left(0\right)}}{2\pi b_{0}}\right],\label{eq:hk_bosonized}
\end{eqnarray}
 where $\phi_{i,c}^{R}\left(y\right),\phi_{i,s}^{R}\left(y\right)$
are bosonic chiral fields which obey the commutation relations $\left[\phi_{i,\nu}^{R}\left(y\right),\phi_{j,\eta}^{R}\left(y^{\prime}\right)\right]=i\pi\text{sign}\left(y-y^{\prime}\right)\delta_{i,j}\delta_{\nu,\eta}$,
and which are related to charge and spin density-fluctuations in the
1D fermionic chains through the relations $\rho_{i}\left(y\right)=-\frac{1}{\pi}\nabla\phi_{i,c}^{R}\left(y\right)$
and $s_{i}\left(y\right)=-\frac{1}{\pi}\nabla\phi_{i,s}^{R}\left(y\right)$,
respectively \cite{giamarchi_book_1d,Kotliar96_Toulouse_points_in_the_generalized_Anderson_model}.
In Eq. (\ref{eq:hf_bosonized}) $v_{F}$ is the Fermi velocity, and
in Eq. (\ref{eq:hk_bosonized}) $\delta_{s}=\tan^{-1}\left(\pi\rho_{0}J_{K}^{z}b_{0}/4\right)$
is the scattering phase-shift associated with the potential $J_{K}^{z}S_{i}^{z}/2$,
$\rho_{0}=\left(2\pi v_{F}\right)^{-1}$ the conduction electron density
of states at the Fermi energy, and $b_{0}$ the lattice parameter
in the fermionic chain. For simplicity we assume these parameters
to be identical for all chains.%
{} We then introduce the (Emery-Kivelson) unitary transformation \cite{emery_kivelson_kondo_review}

\begin{eqnarray}
\mathcal{U} & = & \exp\left[-i\gamma\sum_{i}S_{i}^{z}\phi_{i,s}^{R}\left(0\right)\right],\label{eq:transf_U}
\end{eqnarray}
under which the bosonic field $\nabla\phi_{i,s}^{R}\left(y\right)$
and the spin operator $S_{i}^{+}$ transform as 
\begin{eqnarray}
\mathcal{U}^{\dagger}\nabla\phi_{i,s}^{R}\left(y\right)\mathcal{U} & = & \left[\nabla\phi_{i,s}^{R}\left(y\right)+\delta\left(y\right)2\pi\gamma S_{i}^{z}\right],\label{eq:transformation_1}\\
\mathcal{U}^{\dagger}S_{i}^{+}\mathcal{U} & = & S_{i}^{+}e^{i\gamma\phi_{i,s}^{R}\left(0\right)}.\label{eq:transformation_2}
\end{eqnarray}
 Upon this transformation, the total Hamiltonian of the spin-chain
coupled to the metallic bath, $H=H_{0}+H_{K}+H_{F}$, transforms as
$\tilde{H}=\mathcal{U}^{\dagger}H\mathcal{U}=\tilde{H}_{0}+\tilde{H}_{K}+\tilde{H}_{F}$,
with 
\begin{eqnarray}
\tilde{H}_{0} & = & - \sum_{i}\left\{ J_{H}^{z}S_{i}^{z}S_{i+1}^{z}\right.\nonumber \\
 &  & \left.+\frac{J_{H}^{\perp}}{2}\left[e^{i\gamma\left[\phi_{i,s}^{R}\left(0\right)-\phi_{i+1,s}^{R}\left(0\right)\right]}S_{i}^{+}S_{i+1}^{-}+\text{H.c.}\right]\right\} ,\label{eq:h0_transformed}\\
\tilde{H}_{K} & = & \sum_{i}\left\{ -\frac{2\tilde{\delta}_{s}}{\pi\rho_{0}}S_{i}^{z}\frac{\nabla\phi_{i,s}^{R}\left(0\right)}{\sqrt{2}\pi}\right.\nonumber \\
 &  & \left.+\frac{J_{K}^{\perp}b_{0}}{2}\left[S_{i}^{+}\frac{e^{-i\left(\sqrt{2}-\gamma\right)\phi_{i,s}^{R}\left(0\right)}}{2\pi b_{0}}+\text{H.c.}\right]\right\} ,\\
\tilde{H}_{F} & = & H_{F},\label{eq:hf_transformed}
\end{eqnarray}
 where we have defined $\tilde{\delta}_{s}\equiv\delta_{s}-\pi\gamma/2\sqrt{2}$.
Note that the local bath approximation Eq. (\ref{eq:hf}) is crucial
to implement bosonization along the chains, and to put these ideas
on a clear mathematical framework. It is also interesting to note
that in the transformed representation, the quantum dynamics of the
bath {[}represented by the chiral field $\phi_{i,s}^{R}\left(0\right)${]}
appears explicitly in the Heisenberg term $\sim -J_{H}^{\perp}\left(S_{i}^{+}S_{i+1}^{-}e^{i\gamma\left[\phi_{i,s}^{R}\left(0\right)-\phi_{i+1,s}^{R}\left(0\right)\right]}+\text{H.c.}\right)$
\cite{Lobos12_Dissipative_XY_chain}. Physically, this means that
the Heisenberg interaction is now ``dressed'' by the spin-density
fluctuations of the electron gas. %
{} In the case of an easy-axis spin chain in Ising limit $J_{H}^{z}\gg J_{H}^{\perp}$,
the effect of this term is negligible and can be ignored in a first
approximation.

We now exploit the freedom to choose $\gamma$ and set $\gamma=\sqrt{2}$.
In this case, the Hamiltonian reads

\begin{eqnarray}
\tilde{H} & = & \tilde{H}_{0}+\tilde{H}_{K}+\tilde{H}_{F}\nonumber \\
 & \simeq & \sum_{i}\left[J_{H}^{z}S_{i}^{z}S_{i+1}^{z}-\frac{2\tilde{\delta}_{s}}{\pi\rho_{0}}\frac{\nabla\phi_{i,s}^{R}\left(0\right)}{\sqrt{2}\pi}S_{i}^{z}+\frac{J_{K}^{\perp}}{2\pi}S_{i}^{x}\right]+H_{F},\label{eq:1DDQIM}
\end{eqnarray}
 This Hamiltonian corresponds to the 1D dissipative quantum Ising
model (DQIM), where now the Kondo Hamiltonian $\tilde{H}_{K}$ is
equivalent to the spin-boson model with Ohmic dissipation \cite{Guinea85_Bosonization_of_a_two_level_system,leggett_two_state,Weiss99_Quantum_Dissipative_Systems,Orth08_Dissipative_quantum_Ising_model_in_cold_atoms}
with $\tilde{\delta}_{s}$ related to the dissipative parameter $\alpha$
in the context of macroscopic quantum coherence through $\alpha=\left(2\tilde{\delta}_{s}/\pi\right)^{2}$,
and with the in-plane Kondo interaction playing the role of a magnetic
field along the $x-$axis $h_{x}=-J_{K}^{\perp}/2\pi$. At $T=0$,
the 1D DQIM is known to display a paramagnetic to ferromagnetic quantum
phase transition which is in the universality class of quantum dissipative
systems and whose dynamical critical exponent is $z=2$ \cite{Pankov04_NFL_behavior_and_2D_AFM_fluctuations}.
Note that this is very different with the case of the 1D quantum Ising
chain, which is in the universality class of the 2D classical Ising
model and where $z=1$. The critical properties of this model near
the quantum phase transition have been studied in the context of antiferromagnetic
instabilities of Fermi liquids \cite{sachdev_book,Pankov04_NFL_behavior_and_2D_AFM_fluctuations}
using the framework of the Hertz-Moriya-Millis theory \cite{Hertz76_Quantum_Critical_Phenomena,Moriya73_Critical_phenomena,Millis93_Quantum_critical_points_in_Fermi_systems}
This theory describes critical quantum fluctuations of the order parameter
around the Gaussian fixed point of the theory. The predicted value
of the critical dynamical exponent $z=2$ and has been confirmed numerically
with Monte Carlo simulations \cite{Werner05_Dissipative_quantum_Ising_model_in_1D,Werner05_Quantum_Spin_Chains_with_site_dissipation}.
Physically speaking, such a dynamical exponent implies that the effective
dimensionality of the 1D spin chain coupled to the metallic bath is
$d_{\text{eff}}=1+z=3$, and therefore fluctuations of the order parameter
are expected to be much less important than the case of the non-dissipative
classical or quantum Ising model. This fact supports our MF approximation,
which allows to extend these results to $T>0$.

\ack

AML acknowledges support from DARPA QuEST, JQI-NSFPFC. MAC thanks
Antonio H. Castro Neto for his hospitality at the Graphene Research
Center of the National University of Singapore. We are also grateful
to Piotr Chudzinski for discussions and comments on an early version
of the manuscript.

\bibliographystyle{unsrt} \phantomsection\addcontentsline{toc}{section}{\refname}\bibliographystyle{unsrt}

\end{document}